\documentclass[fleqn,usenatbib]{mnras}

\usepackage{newtxtext,newtxmath}

\usepackage[T1]{fontenc}

\DeclareRobustCommand{\VAN}[3]{#2}
\let\VANthebibliography\thebibliography
\def\thebibliography{\DeclareRobustCommand{\VAN}[3]{##3}\VANthebibliography}

\usepackage{graphicx}	
\usepackage{amsmath}	
\usepackage{amssymb}	

\title[Colour-colour diagrams for Hot Jupiters]{Colour-colour and colour-magnitude diagrams for Hot Jupiters}

\author[G.~Melville et al.]{
G.~Melville$^1$\thanks{E-mail: g.melville@unsw.edu.au}, 
L.~Kedziora-Chudczer$^{1,2}$, J.~Bailey$^1$
\\
\\
$^{1}$School of Physics, UNSW Sydney, NSW, 2052, Australia.\\
$^{2}$Centre for Astrophysics, University of Southern Queensland, Toowoomba, Queensland. 4350. Australia.\\
}

\date{Accepted XXX. Received YYY; in original form ZZZ}

\pubyear{2020}

\begin{document}
\label{firstpage}
\pagerange{\pageref{firstpage}--\pageref{lastpage}}
\maketitle

\begin{abstract}
We use ground-based and space-based eclipse measurements for the near-infrared ($JHK\!s$) bands and Spitzer 3.6 $\mu$m and 4.5 $\mu$m bands to construct colour-colour and colour-magnitude diagrams for hot Jupiters. We compare the results with previous observations of substellar objects and find that hot Jupiters, when corrected for their inflated radii, lie near the black body line and in the same region of the colour magnitude diagrams as brown dwarfs, including low gravity dwarfs that have been previously suggested as exoplanet analogs. We use theoretical emission spectra to investigate the effects of different metallicity, C/O ratios and temperatures on the IR colours. In general we find that while differences in C/O ratio and metallicity do correspond to different locations on these diagrams, the measurement errors are too large to use this method to put strong constraints on the composition of individual objects. However, as a class hot Jupiters cluster around the location expected for solar metallicity and C/O ratio.
\end{abstract}




\section{Introduction}

Hot Jupiters, were among the first discovered exoplanets, because of their size and proximity to their parent star. Characterisation of their atmospheres still remains a challenge due to limited spectroscopic measurements available, which prevents discriminating between the large number of parameters that need to be included in the models. So far significant progress has only been possible for the transiting, short period planets that are most likely tidally locked. Their atmospheres are hot and dynamic, with departures from thermodynamic equilibrium. The expected features in infrared spectra of hot Jupiters include molecular absorption of CO,  CO$_2$, H$_2$O, TiO and possibly CH$_4$, with hydrocarbons such as C$_2$H$_2$, and HCN \citep{2012ApJ...758...36M,2014RSPTA.37230073M}. The actual contribution of each gas varies significantly depending on the atmospheric temperature and planetary composition, including the metallicity and carbon to oxygen ratio in the atmosphere. 

Atmospheres are further complicated by the likely presence of dust, aerosols and clouds, and in some of the cooler of the hot Jupiters these can obscure some spectral features. Likewise, strong optical or UV absorbers in the upper layers of the atmosphere can cause thermal inversion \citep{2010plsc.book.....D} that would cause spectral absorption features to appear in emission. Although a previous analysis of secondary-eclipse spectra of nine hot Jupiters \citep{2013ApJ...779....3L} suggests little evidence for thermal inversions over a wide range of effective temperatures (with the exception of one of the hottest planets, HD209458b). 
The nature of the possible responsible absorbers is as yet unknown in this high-temperature regime. It has been proposed that highly irradiated planets are warm enough for significant amounts of gaseous TiO and VO to exist in the upper atmosphere, providing the necessary opacity to generate a temperature inversion \citep{2003ApJ...594.1011H,2006ApJ...652..746F, 2008ApJ...683.1104F,2008ApJ...678.1436B}. No observational evidence for TiO and VO absorption has been found in the past, until recent detections of TiO in observations of WASP-33b \citep{2017AJ....154..221N} and WASP-19b \citep{2017Natur.549..238S}. 

The atmospheres of highly irradiated, tidally locked planets show large temperature variations between dayside and nightsides, determined by atmosphere’s ability to transport the energy incident upon the substellar point to the nightside. The phase curve measurements at 8$\mu$m  for hot Jupiter, HD 189733b \citep{2007Natur.447..183K}, show a minimum and maximum brightness temperature of 973$\pm$33 K and 1212$\pm$11 K respectively, which indicates a relatively efficient transfer of energy from the dayside to the nightside. This has contrasted with the most irradiated hot Jupiters, such as, for example, WASP-18b \citep{2013Icar..226.1719I}.

Despite obvious differences in irradiation levels, hot Jupiters have similar equilibrium effective temperatures to late type brown dwarfs, which show characteristic distributions on the colour-magnitude and colour-colour diagrams that indicate changes in opacity of their atmospheres \citep{2015MNRAS.450.2279T}. It is desirable to compare where hot Jupiters are placed in such diagrams in order to understand the differences in their atmospheric composition and opacities \citep{triaud14a,triaud14b} . \citet{zhou15} plot colours $K\!s-$[3.6] versus $K\!s-$[4.5] derived from the 2.15$\mu$m $K\!s$ band observed with IRIS2 at the Anglo Australian Telescope (AAT) and from Spitzer 3.6$\mu$m and 4.5$\mu$m measurements for the hot Jupiters and brown dwarfs. The systematic offset between two types of objects is visible despite the large scatter in colour for the exoplanets. \citet{triaud14a} reports a similar scatter where diversity of colour for hot Jupiters increases with decreasing luminosity. This variety of colour reflects the large range in temperatures of irradiated planets.

Since these studies were done further data on the Spitzer 3.6$\mu$m, 4.5$\mu$m and $K\!s$ band eclipse depths have become available, and in addition it has become possible to use WFC3 at the Hubble Space Telescope to obtain measures of the $J$ and $H$ band eclipse depths to supplement the very small number of ground-based measurements at these wavelengths. In addition accurate GAIA DR2 parallaxes are now available for most objects \citep{gaia18}.

In this paper we generate a simplified model spectra of hot Jupiters within a range of atmospheric parameters, and place them on the colour-magnitude and colour-colour diagrams to understand which parameters generate scatter in the colours. We also aim to identify trends, which show in the diagrams with changes in particular parameters of the models. In Section~\ref{section2} we describe our derivation of the observed near-infrared magnitudes and colours for hot Jupiters. In Section~\ref{section3} we introduce the range of parameters that are used in our models, and describe our modelling technique using the VSTAR radiative transfer package. In Section~\ref{section4} we present and discuss the colour-magnitude diagrams and in Section~\ref{section5} the colour-colour diagrams. In Section~\ref{section6} we discuss the trends in the diagrams that are visible when the parameters, such as C/O, metallicity, temperature and gravity are varied. We also compare the models with hot-Jupiters measurements using our IRIS2 observations and other data previously published.  

\section{Observed magnitudes and Colours of hot Jupiters}
\label{section2}

To obtain the absolute magnitudes of a hot Jupiter, we need, for each required band, the secondary eclipse depth that measures the flux of the planet as a fraction of that from the star and also the apparent magnitude of the star. In addition we need the parallax of the object.

Table~\ref{tab:3} lists the eclipse depths for 28 systems which are all hot Jupiters except for the hot brown dwarf KELT-1b. We have included all systems that have both published Spitzer 3.6$\mu$m and 4.5$\mu$m measurements and ground-based $K\!s$ band measurements. We also added the two well-studied systems HD 189733b and HD 209458b which lack a $K\!s$ band measurement but are two of the coolest hot Jupiters included. There are many more systems that have Spitzer measurements \citep[e.g.][]{garhart19} but are excluded from our sample as they have no Ks band measurements.

The $K\!s$ band eclipse depths are from published ground-based near-IR photometry and include our own measurements with the IRIS2 instrument on the Anglo-Australian Telescope \citep{2014MNRAS.445.2746Z,zhou15,chudczer19}. These ground-based observations are challenging and there have been suggestions that the results should be treated with caution and the uncertainties may be underestimated  \citep[e.g][]{demooij11}. However, note that in Table~\ref{tab:3} there are 11 systems that have two $K\!s$ band eclipse depth measurements from independent groups and these are in good agreement given the quoted errors. As there is currently no space instrument making such measurements at this wavelength the ground-based measurements are the only way of obtaining these eclipse depths.

For the $J$ and $H$ bands there are a few ground-based eclipse depth measurements but these wavelengths are more difficult due to the usually smaller eclipse depths, and the errors are generally large. However, the $J$ band and part of the $H$ band are covered by the HST/WFC3 instrument which produces spectra over the range 1.1 -- 1.7$\mu$m. This instrument has been used to obtain emission spectra (by measuring the eclipse depth) for a number of hot Jupiters. We used the published spectra to obtain $J$ and $H$ eclipse depths by summing the spectral bins that fall within the $J$ and $H$ filter passbands as defined by \citet{tokunaga02}. These measurements are indicated by an H in Table~\ref{tab:3}. It should be noted that while the $J$ band is fully covered by WFC3, the coverage of the $H$ band which extends to 1.78$\mu$m is only partial. Therefore if the eclipse depth is increasing with wavelength it may be underestimated by this method. The $J$ band measurements from HST/WFC3 should therefore be regarded as more reliable than the $H$ band measurements.

\begin{table*}
{
\caption{\label{tab:3} Eclipse depths (in per cent) for hot Jupiters. For the $J$ and $H$ band G indicates ground-based measurements, and H indicates measurements derived from published HST/WFC3 spectra.}
\begin{tabular}{lllllll}
    \hline \hline
   Name & $J$ & $H$ & $K\!s$ & [3.6] & [4.5] & Reference 
 \\ 
   \\
   \hline
WASP-3b	&       	&         &	  0.181$\pm{0.020}$        &0.209$^{+0.040}_{-0.028}$	       &0.282$\pm{0.012}$	         &1, 2	\\
        &           &         &   0.193$\pm0.014$      &     &      &   3 \\
WASP-4b	&	 0.057$\pm$0.008H   &   0.086$\pm$0.010H      &	  0.185$^{+0.014}_{-0.013}$         & 0.319$\pm{0.031}$ & 0.343$\pm{0.027}$   &	4, 5, 6\\
 	&	        &         &	  0.16$\pm{0.04}$         &    &    &  7	\\
WASP-5b	& 0.168$\pm{0.05}$G	&         &	0.269$\pm{0.062}$	   &  0.197$\pm{0.028}$ & 0.237$\pm{0.024}$   &8, 9\\
   &            &         &   0.20$\pm$0.02           &    &    &  7   \\
WASP-10b &  &  &   0.137$^{+0.013}_{-0.019}$  &  0.100$\pm0.011$ & 0.146 $\pm0.016$ &  10, 11 \\    
WASP-12b& 0.119$\pm{0.017}$H &0.192$\pm{0.017}$H &  0.296$\pm{0.014}$    & 0.421$\pm{0.011}$ & 0.428$\pm{0.012}$	 & 3, 12, 13 \\
 & 0.139$\pm{0.027}$G  &0.191$\pm{0.016}$G &  0.323$\pm$0.011  &  & 	 &    14, 15, 16 \\
WASP-14b&    &     &    0.172$\pm$0.025    &   0.182$\pm$0.007&   0.216$\pm$0.009 &  16, 17\\
WASP-18b& 0.0847$\pm{0.0026}$H	& 0.1014$\pm{0.0038}$H        &	0.13$\pm{0.03}$   & 0.304$\pm{0.017}$  & 0.379$\pm{0.018}$	 &7, 18, 19, 20 \\
    &     &    &   0.149$\pm0.014$     &   &   &    21  \\
WASP-19b& 0.171$^{+0.075}_{-0.073}$G	&0.276$\pm{0.044}$G &	0.366$\pm{0.072}$   & 0.483 $\pm{0.025}$  & 0.572$\pm{0.03}$  &	 22, 23, 23a\\
&  & 	&0.287$\pm{0.02}$ &	      &  	 & 24  \\
WASP-33b&  0.115$\pm{0.007}$H     	&   0.119$\pm{0.010}$H      &	0.27$\pm{0.04}$	       & 0.26$\pm{0.05}$  & 0.41$\pm{0.02}$	 &25, 26\\
   &      &    &   0.244$^{+0.027}_{-0.020}  $   &    &    &    27   \\
WASP-43b      &  0.045$\pm$0.002H   &   0.060$\pm$0.002H   &    0.181$\pm$0.027    & 0.345$\pm$0.013  &  0.382$\pm$0.015  &   24, 28, 29  \\
             &    &   0.103$\pm$0.017 &   0.194$\pm$0.029  &   &     &    30   \\
WASP-46b&  	&    &	0.26$^{+0.05}_{-0.03}$  &  0.136$\pm$0.070 & 0.445$\pm$0.059 	& 7, 17 \\
        &    0.129$\pm$0.055G     &   0.194$\pm$0.078G      & 0.253$^{+0.063}_{-0.060}$   &      &     & 31  \\
WASP-48b&        &   0.047$\pm$0.016G  &  0.109$\pm$0.027  & 0.176$\pm$0.013 & 0.214$\pm$ 0.020 & 32 \\        
     &     &     &  0.136$\pm$0.014   &     &    &    33   \\
WASP-103b&  0.126$\pm$0.007H  & 0.158$\pm$0.008H  &   0.357$^{+0.040}_{-0.035}$   &  0.335$\pm$0.022 & 0.471$\pm$0.036& 17, 34, 35 \\
CoRoT-1b&       	&  0.145$\pm{0.049}$       &	 0.336$\pm{0.042}$         &0.415$\pm{0.042}$	       &	0.482$\pm{0.042}$	     &   36, 37, 38\\
CoRoT-2b&	0.032$\pm$0.004H    &  0.051$\pm$0.004H    &	0.16$\pm{0.09}$  & 0.355$\pm{0.02}$  &	0.51$\pm{0.041}$	&37, 39, 40\\
KELT-1b	&	        &  0.149$\pm{0.009}$G       &	0.16$\pm{0.018}$   &	0.195$\pm{0.01}$&
0.2$\pm{0.012}$ &	3, 41, 42\\
KELT-2Ab   & &    &   0.118$\pm$0.017   &    0.0739$\pm$0.0043  &  0.0761$\pm$0.0053 & 16, 17 \\
HAT-P-1b&	        &         &	0.109$\pm{0.025}$   &	0.08$\pm{0.008}$&	0.135$\pm{0.022}$&	43, 44\\
HAT-P-23b&          &    &   0.234$\pm$0.046   &  0.248$\pm$0.019 & 0.309$\pm$0.026 &  32 \\
HAT-P-32b&  0.035$\pm{0.010}$H     	& 0.090$\pm{0.033}$G&	0.178$\pm{0.057}$   &	0.364$\pm{0.016}$&	0.438$\pm{0.02}$ &	45, 46\\
HAT-P-33b&     &     &    0.153$\pm$0.022   &    0.166$\pm$0.013   &   0.190$\pm$0.021 &  16, 17 \\
TReS-2b	&       	&         &	0.062$^{+0.013}_{-0.011}$   &	0.127$\pm{0.021}$&	0.23$\pm{0.024}$ &	47, 48\\
TReS-3b	&  0.038$\pm$0.011H	&   0.064$\pm$0.010H  &	 0.133$^{+0.018}_{-0.016}$    &	0.346$\pm{0.035}$&	0.372$\pm{0.054}$	&49, 50, 6 \\
Kepler-5b&   &    &   0.080$\pm$0.025  &   0.103$\pm$0.017   &   0.107$\pm$0.015 &   16, 51 \\
Kepler-13Ab	&  0.051$\pm{0.007}$H     	& 0.109$\pm{0.008}$H        &	0.122$\pm{0.051}$   &0.156	 $\pm{0.031}$	&	0.222	 $\pm{0.023}$ &  52, 53\\
Qatar-1b&       &    &   0.136$\pm$0.034  &   0.151$\pm$0.045  &  0.291$\pm$0.42  &  3, 17 \\
        &    &   &   0.196$^{+0.071}_{-0.051}$   &   &   &  54   \\
HD 189733b &	0.0076$\pm$0.0040H &  0.0152$\pm$0.0040H  &  &	0.147$\pm{0.004}$&	0.179$\pm{0.004}$  &	 55, 56\\
HD 209458b&  0.0109$\pm$0.0009H    	&   0.0148$\pm$0.0010H   &   &	0.094$\pm{0.009}$&	0.213$\pm{0.015}$	&57, 58\\

    \hline
    
\end{tabular}}
  \newline
  \newline
  \newline
  References:
  1) \citet{zhao12}; 2) \citet{rostron14}; 3) \citet{croll15}; 4) \citet{beerer11}; 5) \citet{caceres11}; 6) \citet{ranjan14}; 7) \citet{zhou15}; 8) \citet{chen14}; 9) \citet{baskin14}; 10) \citet{cruz15};  11) \citet{kammer15}; 12) \citet{stevenson14}; 13) \citet{swain13}; 14) \citet{croll11}; 15) \citet{crossfield12}; 16) \citet{martioli18}; 17) \citet{garhart19}; 18) \citet{nymeyer11}; 19) \citet{maxted13}; 20) \citet{arcangeli18}; 21) \citet{chudczer19}; 22) \citet{gibson10}; 23) \citet{anderson13}; 23a) \citet{lendl13}; 24) \citet{2014MNRAS.445.2746Z}; 25) \citet{haynes15}; 26) \citet{deming12}; 27) \citet{demooij13}; 28) \citet{stevenson15}; 29) \citet{blecic14}; 30)  \citet{wang13};  31) \citet{chen14b}; 32) \citet{orourke14}; 33) \citet{clark18}; 34) \citet{delrez18}; 35) \citet{cartier17}; 36) \citet{rogers09}; 37) \citet{deming11}; 38) \citet{zhao12b}; 39) \citet{alonso10}; 40) \citet{wilkins14};  41) \citet{beatty14}   42); \citet{beatty17a}; 43) \citet{demooij11}; 44) \citet{todorov10};  45) \citet{zhao14}; 46) \citet{nikolov18}; 47) \citet{croll10}; 48) \citet{odonovan10}; 49) \citet{fressin10}; 50) \citet{croll10b};  51) \citet{desert11}; 52) \citet{shporer14}; 53) \citet{beatty17b};  54) \citet{cruz16}; 55) \citet{knutson12}; 56) \citet{crouzet14}; 57) \citet{knutson08}; 58) \citet{line16}. 
\end{table*}

\begin{table*}
{
\caption{\label{tab:4} Absolute magnitudes for hot Jupiters derived from the data in Table~\ref{tab:3} on the zero Vega magnitude convention as described in Section~\ref{section2}.}
\begin{tabular}{lrrrrr}
    \hline \hline
   Name & M$_{J}$ & M$_{H}$ & M$_{K\!s}$ & M$_{[3.6]}$ & M$_{[4.5]}$ 
 \\ 
   \\
   \hline
WASP-3b	&       	&         &	  9.34$\pm{0.07}$        &9.23$\pm{0.18}$	       &8.89$\pm{0.08}$	   \\
WASP-4b	& 12.14$\pm$0.16 & 11.36$\pm$0.13 &	10.45$\pm{0.08}$ & 9.80$\pm{0.12}$ & 9.74$\pm{0.10}$   \\
WASP-5b	& 10.42$\pm{0.33}$	&         &	9.84$\pm{0.10}$	  &  9.83$\pm{0.17}$ & 9.68$\pm{0.13}$  \\
WASP-10b	&	        &         &	 11.39$\pm$0.09   & 11.67$\pm{0.12}$  & 11.33$\pm{0.12}$    \\
WASP-12b& 9.55$\pm{0.12}$ &8.85$\pm{0.07}$ &  8.27$\pm{0.04}$    & 7.87$\pm{0.05}$ & 7.84$\pm{0.05}$\\
WASP-14b&            &   &    9.47$\pm$0.16  &   9.38$\pm$0.06  &  9.18$\pm$0.06  \\
WASP-18b& 10.62$\pm{0.14}$	& 10.25$\pm$0.07 &	9.75$\pm{0.10}$   & 8.92$\pm{0.08}$  & 8.70$\pm{0.07}$	\\
WASP-19b& 10.67$\pm$0.50  	&  9.84$\pm{0.18}$ &	9.65$\pm{0.07}$   & 9.08$\pm{0.07}$  & 8.93$\pm{0.07}$ \\
WASP-33b&  9.49$\pm$0.07  &  9.39$\pm$0.09  &	8.53$\pm{0.09}$   & 8.45$\pm{0.22}$  & 7.97$\pm{0.08}$	\\
WASP-43b&   13.67$\pm$0.05 & 12.17$\pm$0.18 & 11.39$\pm$0.12 & 10.59$\pm$0.08 & 10.56$\pm$0.07  \\
WASP-46b&   11.08$\pm{0.50}$ &  10.35$\pm{0.46}$  &	9.98$\pm{0.15}$  & 10.62$\pm$0.62  & 9.36$\pm$0.15   \\
WASP-48b&     &  10.45$\pm$0.39   &  9.27$\pm$0.10  &  8.91$\pm$0.10 &  8.72$\pm$0.12  \\
WASP-103b&  9.99$\pm$0.07 & 9.50$\pm$0.06 &	8.52$\pm$0.12  &  8.54$\pm{0.08}$ & 8.20$\pm{0.09}$  \\
CoRoT-1b&  	&  9.78$\pm{0.38}$ & 8.80$\pm{0.14}$  &8.54$\pm{0.16}$  &	8.38$\pm{0.13}$	   \\
CoRoT-2b&12.86$\pm$0.14  & 12.01$\pm$0.09 &	10.64$\pm{0.69}$  & 9.76$\pm{0.09}$  &	9.38$\pm{0.12}$	\\
KELT-1b	&	    & 9.41$\pm$0.10  &	9.27$\pm{0.12}$  &	9.00$\pm{0.08}$&	9.08$\pm{0.09}$ \\
KELT-2Ab&     &   &  9.15$\pm$0.06  &  9.64$\pm$0.07  &  9.56$\pm$0.08   \\
HAT-P-1b&	        &         &	10.25$\pm{0.25}$   &	10.60$\pm{0.13}$&	10.01$\pm{0.19}$\\
HAT-P-23b&       	&         &	 9.54$\pm$0.22  &	9.50$\pm{0.11}$&	9.16$\pm{0.11}$ \\
HAT-P-32b& 11.57$\pm$0.32 &  10.32$\pm{0.42}$  & 9.54$\pm{0.36}$ & 8.68$\pm{0.05}$&	8.48$\pm{0.05}$ \\
HAT-P-33b&   &   &  9.03$\pm$0.16  &  8.93$\pm$0.09  &   8.80$\pm$0.12  \\
TReS-2b	&       	&         &	11.19$\pm{0.21}$   & 10.34$\pm{0.19}$&	9.71$\pm{0.12}$ \\
TReS-3b	& 12.73$\pm$0.32 &  11.80$\pm$0.17 & 10.96$\pm{0.14}$  &	9.87$\pm{0.13}$& 9.84$\pm{0.17}$\\
Kepler-5b&    &   &   9.70$\pm$0.35  &  9.34$\pm$0.18 &  9.35$\pm$0.15     \\
Kepler-13Ab&  9.82$\pm$0.14 & 8.98$\pm$0.08 & 8.83$\pm$0.48 &	8.52$\pm{0.22}$&	8.16$\pm{0.11}$\\
Qatar-1b&	     	&         &	 11.11$\pm$0.22 &	11.02$\pm{0.33}$& 10.38$\pm{0.16}$\\
HD 189733b&  14.89$\pm$0.64 	&  13.65$\pm$0.29  &    &	11.05$\pm{0.07}$&	10.92$\pm{0.07}$\\
HD 209458b &	 13.07$\pm$0.09  &  12.52$\pm$0.08  &       &	10.40$\pm{0.11}$&	9.56$\pm{0.04}$\\
    \hline
    
\end{tabular}}
  \newline
  \newline
  \newline
\end{table*}

The apparent magnitudes for the host stars have, where possible, been taken from the compilation by \citet{triaud14b}. For eight stars not included in that study we used the $JHKs$ magnitudes from the 2MASS all-sky catalog \citep{cutri03}. The [3.6] and [4.5] magnitudes for these objects were taken from the W1 and W2 magnitudes in the ALLWISE source catalog \citep{wright10,mainzer11}. \citet{triaud14b} show that there is good agreement between the W1 and W2 WISE bands and the [3.6] and [4.5] bands as measured with Spitzer. Two of the new objects (Kepler-13A and KELT-2A) are in blended binaries and we have corrected the magnitudes using the dilution factors given in \cite{shporer14}, \cite{martioli18} and \cite{garhart19}. We note that since the dilution factor is applied to both the eclipse depth, and to the host star magnitude, it cancels out in the eventual determination of the planet absolute magnitude and therefore does not contribute to the uncertainty in that determination.

In the analyses of \cite{triaud14a} and \cite{triaud14b} few objects had well determined parallaxes and photometric distances were used for many objects. We can now replace these with GAIA DR2 \citep{gaia18} parallaxes which are available for most of our objects and have typical precisions of $\sim$1\% or better. This means that the error in the parallax is negligible compared with the errors in the apparent magnitudes and eclipse depths. However for WASP-103 the GAIA-DR2 parallax has a very large error (15\%) and we used instead the distance of 470$\pm$35 pc from \citet{gillon14}.

Our final absolute magntiudes derived from the eclipse depths and other information described above are given in Table~\ref{tab:4}. Where there was more than one measurement of eclipse depth in Table~\ref{tab:3} a weighted combination was used. Of the 28 systems in the table, 26 have $K\!s$ magnitudes, 17 have $H$ magnitudes and 14 have $J$ magnitudes. The results from our study are mostly in good agreement with those in \cite{triaud14b}, but our errors are generally smaller, due largely to the improved parallaxes, and that study only included one $J$ magnitude and 4 $H$ magnitudes.

\section{Hot-Jupiter Models}
\label{section3}

We used the VSTAR (Versatile Software for Transfer of Atmospheric Radiation), an atmospheric radiative transfer software package, \citep{2012MNRAS.419.1913B} to construct our models of hot Jupiter spectra. This code is designed for application to a wide range of atmospheres, including those of the Earth, planets, brown dwarfs and cool stars. VSTAR solves the radiative transfer problem in a plane-parallel atmosphere by combining a line-by-line treatment of the molecular absorption with a full multiple-scattering solution of the radiative transfer equation. Our models are for a planet around the sun-like star (G2V). The mass and equatorial radius of the planet are assumed to be  identical to Jupiter (71492 km) giving a surface gravity of g=24.8 ms$^{-2}$. Previous applications of VSTAR to exoplanet atmospheres are described in \citet{2014MNRAS.445.2746Z}, \citet{zhou15}, \citet{chudczer19} and \citet{bailey18}.

The models have been calculated with a range of metallicity and C/O ratios with the aim of investigating whether the observed colours of a hot Jupiter can be used to constrain the composition of the planet's atmosphere.

\begin{figure}
    \centering
    \includegraphics[width=8cm]{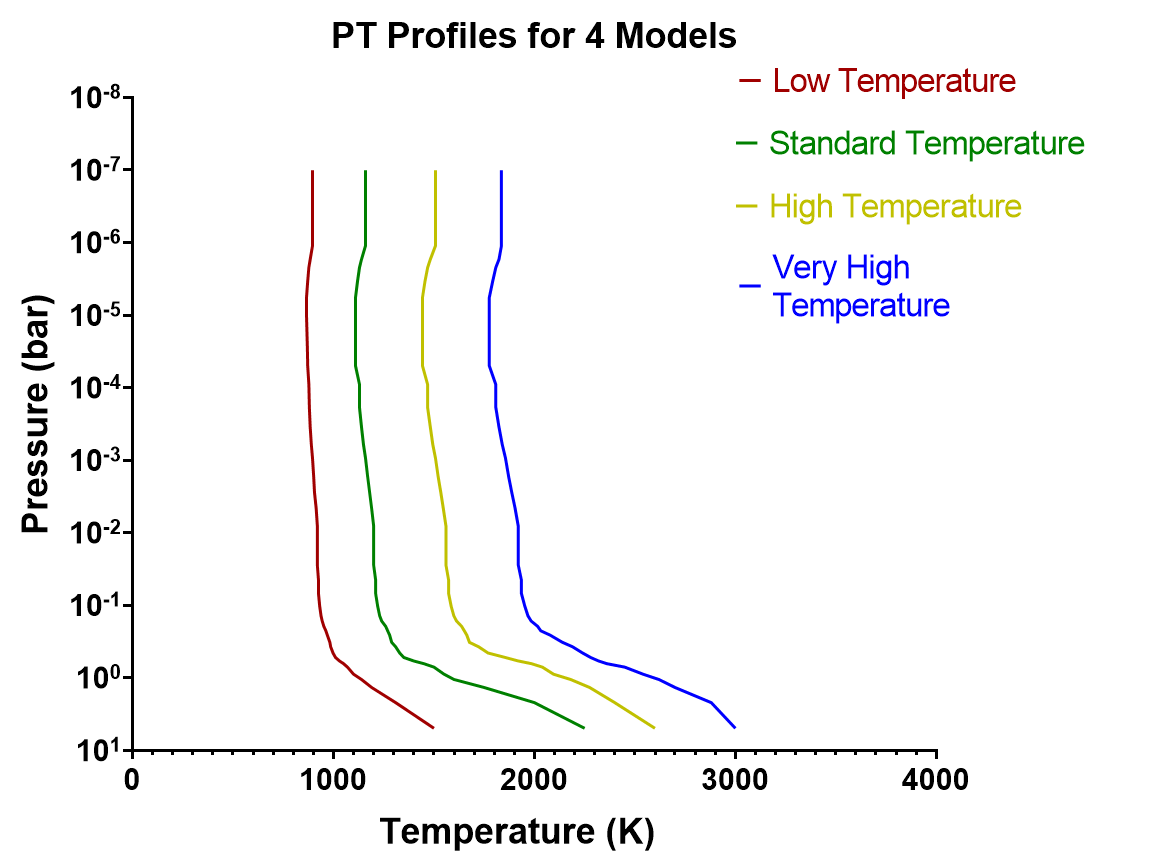}
     \caption{P-T profiles used in our grid of models. Medium temperature (green) profile is referred to as the Standard Temperature P-T profile.} 
    \label{fig:pt}
\end{figure}

The atmospheric pressure-temperature structures are shown in Fig.~\ref{fig:pt} and are adapted from the profile we have used for modelling HD~189733b \citep{bailey18}, which will be referred to as the standard P-T profile (ST). The low (LT), high (HT) and very high (VHT) temperature models were derived from this as specified in Table~\ref{tab:1} and translate to different orbital distances of the planet from the star.
 
 The molecular composition of the atmosphere is calculated under the assumption of chemical equilibrium using the methods described in \citep{2012MNRAS.419.1913B}. We do not include non-equilibrium processes such as photochemistry which is mostly important at higher levels in the atmosphere \citep{2014prpl.conf..739M} than are important for the emission spectrum. We calculated models for each of the four P-T profiles from Fig.~\ref{fig:pt} for different combination of metallicity and C/O ratio as listed in Table~\ref{tab:2}. We used the Ionization and Chemical Equilibrium (ICE) package of VSTAR to determine the equilibrium abundances, expressed in terms of mixing ratios, in every atmospheric layer using a database of 143 compounds in gaseous and condensed phases. 

In the next step multiple-scattering, radiative transfer calculations were performed for wavelengths from 0.324 $\mu$m to 28.6 $\mu$m giving a spectrum with  304501 points (0.1 cm$^{-1}$ resolution). The opacities in this final spectrum were derived by using a comprehensive database of molecular spectral lines as described in \citet{2012MNRAS.419.1913B}. We include spectral lines of the following  species: H$_{2}$O, CO, CH$_{4}$, CaH, MgH, CrH, FeH, TiH, Na, K, Rb, Cs, HCN, C$_{2}$H$_{2}$, CO$_{2}$., TiO and VO. In addition we included Rayleigh scattering by H$_{2}$, He, H, the opacities from the collisionally-induced absorption due to H$_{2}$–H$_{2}$ and H$_{2}$–He \citep{1998STIN...9953342B}, and the free-free and bound-free absorption from H, H$^{-}$ and H$_{2}^{-}$. These calculations were performed at each wavelength in the spectral range considered for each of 42 pressure layers between  10$^{-7}$ and 5 bars. 

The VSTAR models are are used to derive the wavelength dependent flux at the top of the exoplanet atmosphere. 
To derive the magnitudes and colours from these modelled spectra, we used the average intensities over the bands with ranges corresponding to the bands of the IRAC/Spitzer camera and the $J$, $H$, and $K\!s$ bands of the IRIS2/AAT imager. Absolute magnitudes were derived according to:

\begin{equation}
    M = 2.5 \log(\frac{f_{Vega}}{f_{*}}),
\label{eq:1}
\end{equation}

where fluxes were standarised at a distance of 10 parsecs and absolute magnitudes calculated using the Vega zero point appropriate for each spectral band \citep{1998A&A...333..231B}. Colours used in our plots were calculated by taking a difference of magnitudes derived from Equation~\ref{eq:1}. 

 \begin{table}
{\scriptsize
\caption{\label{tab:1} PT profiles considered in our models}
\begin{tabular}{lrrl}
    \hline \hline
   Name  & \multicolumn{2}{c}{Temperatures} & Orbital radius \\ 
        &  TOA     &   BOA    & of the planet\\
      Units   & \textit{K} &\textit{K} &\textit{AU} \\
    \hline
     \bf{Low} (LT) & 896 & 1500 & 0.12\\ 
     \bf{Standard} (ST) & 1160 & 2250 & 0.07\\ 
     \bf{High} (HT) & 1508 & 2600 & 0.04\\    
     \bf{Very High} (VHT) & 1836 & 3000 & 0.027\\ 
    \hline
\end{tabular}}
\end{table}

\begin{table}
{\scriptsize
\caption{\label{tab:2} List of parameters and their values tested  }
\begin{tabular}{ccccccc}
    \hline \hline
     Name & \multicolumn{5}{c}{List of values}\\
 \hline
      Z defined as $[Fe/H]$  & -0.6 & -0.4&0 &0.48 & 0.6 \\
    VSTAR Z$\times$S$\odot$ &0.25 &0.4 &1 &3&4  \\ 
   
    C/O ratio &  0.1 &0.4& 0.7 & 1.6 &2.0 \\
   
    \hline
\end{tabular}}
\end{table}

\begin{figure*}
    \centering
    \includegraphics[width=15cm]{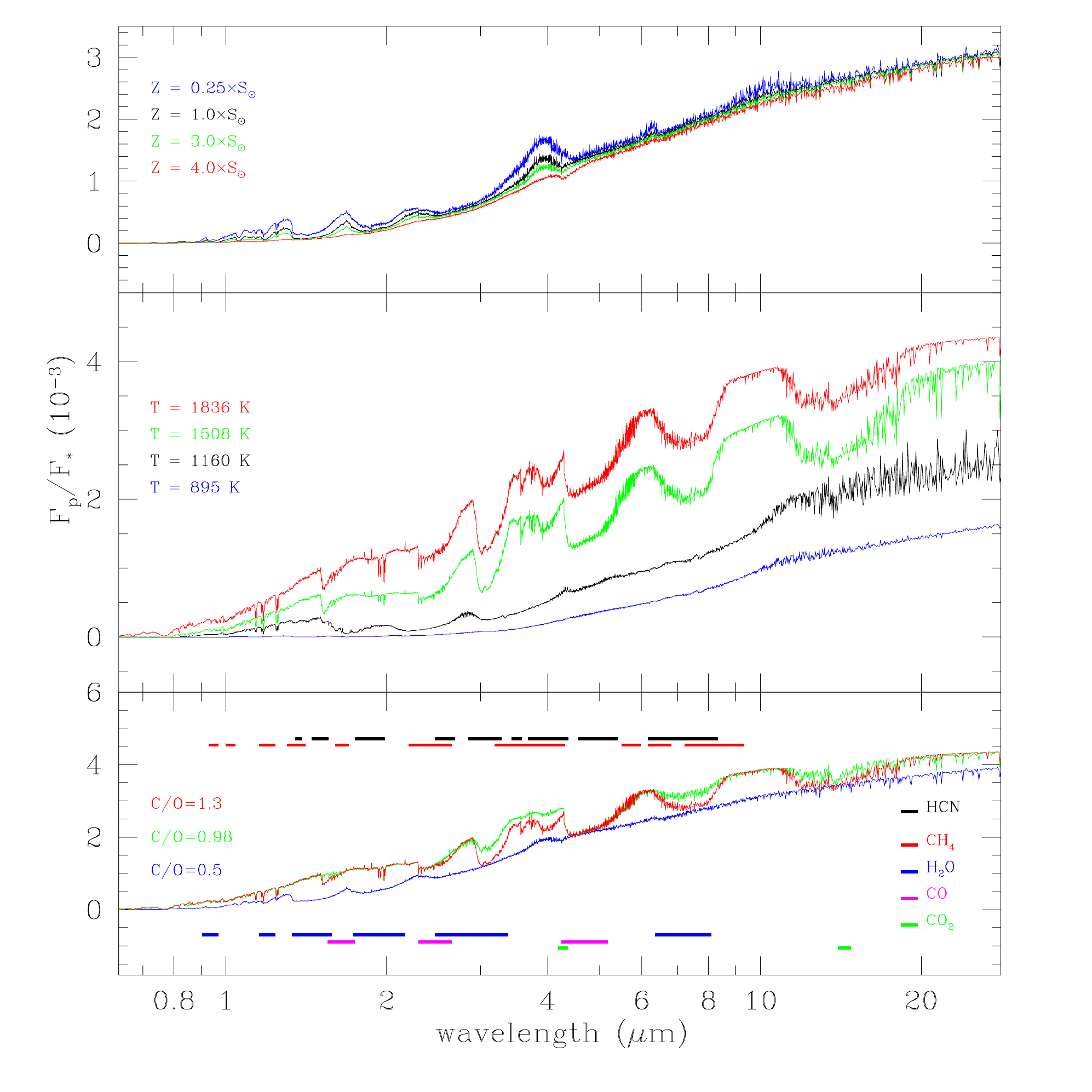}
     \caption{Emission spectra for the selected parameters. Top panel: different metallicity ratios for the VHT models. Middle panel shows the difference in spectra at four considered P-T profiles assuming solar metalicity and C/O ratio of 1.6. Bottom panel shows the differences in the VHT spectra under solar metallicity and changing C/O ratio.} 
    \label{fig:spectra}
\end{figure*}

 Fig.~\ref{fig:spectra} shows examples of the model spectra when one of the parameters is varied. They are smoothed to the resolution of R $\sim1000$ at $K\!s$ band. 
 In the top panel we varied the metallicity for the very high temperature (VHT) models under the same Solar value of C/O ratio. In middle panel the models with different temperature profiles for the same Solar metallicity and C/O ratio, are plotted. Finally in the bottom panel we present the results of changes in the C/O ratio for models with VHT profile and with assumed Solar metallicity. 

 The graphs in Fig.~\ref{fig:spectra} show the importance these different parameters have on absorption features due to different molecules in the planetary atmosphere. For example the most significant absorption in the range of temperatures considered, is due to H$_{2}$O in near- and mid-infrared. When the temperature is reduced, the scale height of the atmosphere is smaller which leads to shallower features in the spectrum. Similarly the increase in metallicity has the same effect on the scale height. Variation in the C/O ratio has more complex effect on the spectra and it shows most strongly around C/O$=1$ where the dominance of absorption from oxygen-rich species like H$_{2}$O is being replaced with increasing contribution of carbon-rich species such as CH$_4$, HCN and H$_{2}$C$_{2}$. 
 
 Of course the relative contributions of these molecules depend on the temperature and composition of the atmosphere \citep{2005ApJ...632.1122S,2011Natur.469...64M,2012ApJ...745...77K,2011ApJ...737...15M}. For example the high temperature spectra will show a significant continual decrease in atmospheric H$_{2}$O with increasing metallicity. This is because generally an enhancement of metallicity in H/He-dominated atmospheres of gas giant planets favours an increase in the abundances of molecules that contain multiple heavy atoms, i.e., CO and N$_{2}$ are favoured over the lighter CH$_{4}$, NH$_{3}$ and H$_{2}$O. Molecules with more than two heavy atoms are even more favoured, as occurs in the case of CO$_{2}$, whose abundance increases as the square of metallicity \citep{2009ApJ...701L..20Z,2009arXiv0911.0728Z}. Their simulations indicate that H$_{2}$O forms more easily in planetary atmospheres when both the metallicity and the C/O ratio are relatively low. 

\citet{2009ApJ...699.1487S} suggested that TiO and VO may be depleted from the upper atmospheres due to gravitational settling coupled with condensation. Since they are significantly heavier than molecular hydrogen, the primary constituent of a hot Jupiter atmosphere, this would lead to a requirement of a significant turbulent diffusion or large-scale advection to maintain a high abundance of heavy species at high altitudes. Cross-planet circulation dominated by strong equatorial jets may also deplete TiO and VO from the planetary dayside. However, HCN, TiO and H$_{2}$C$_{2}$ are very prominent in our HT and VHT models, that is, above 1500 K where they are in the gas phase. 

\begin{figure*}
    \centering
    \includegraphics[width=8.8cm]{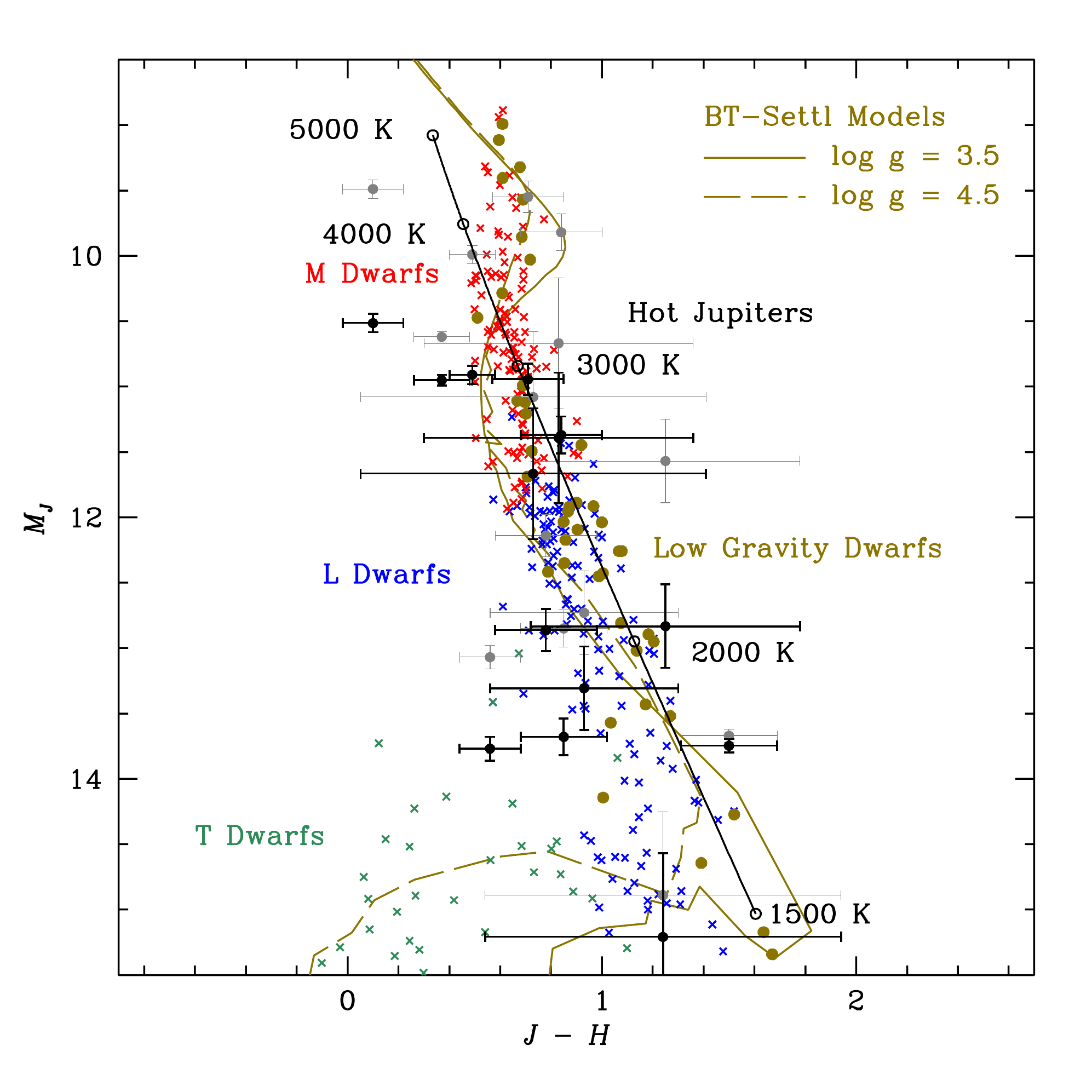}
    \includegraphics[width=8.8cm]{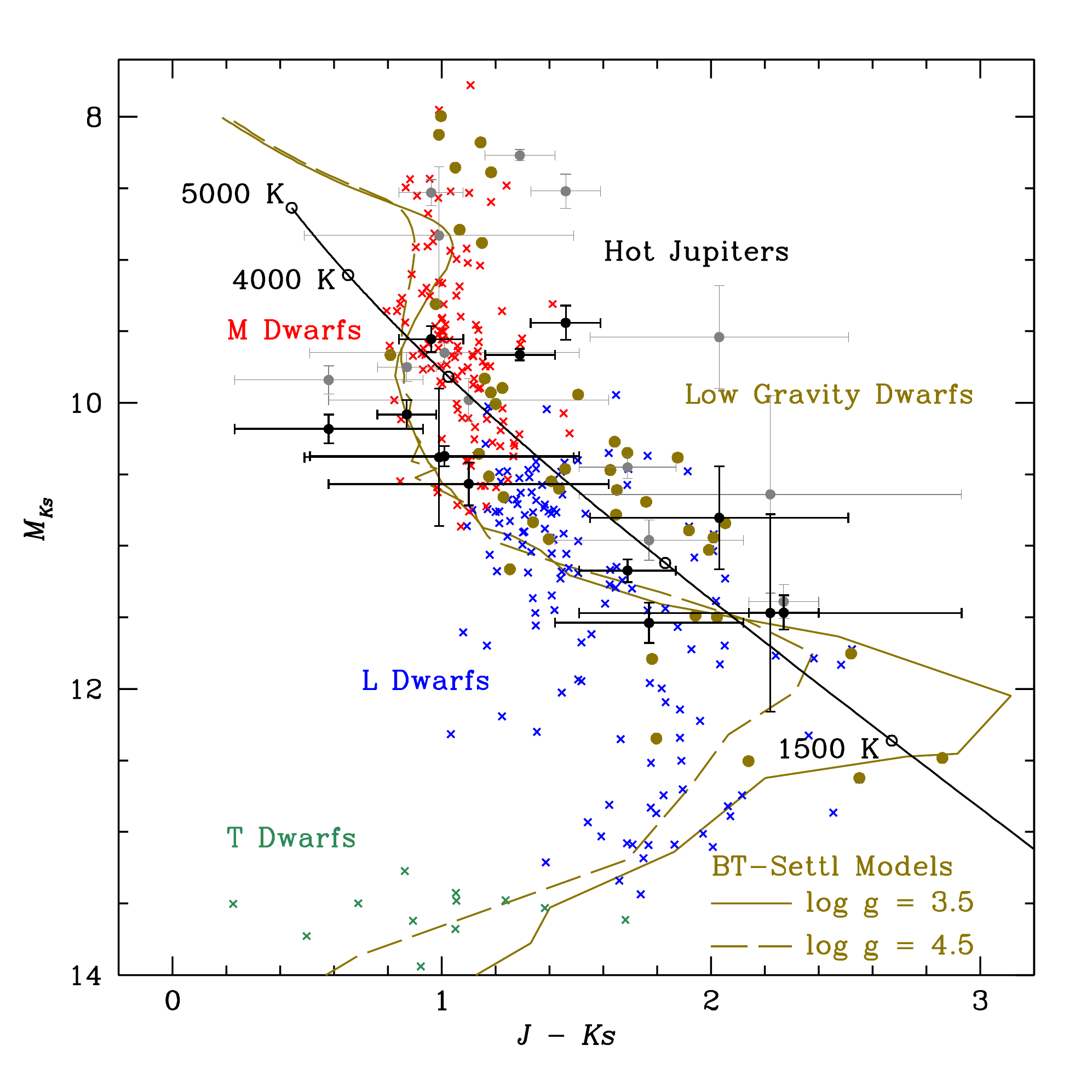}
     \includegraphics[width=8.8cm]{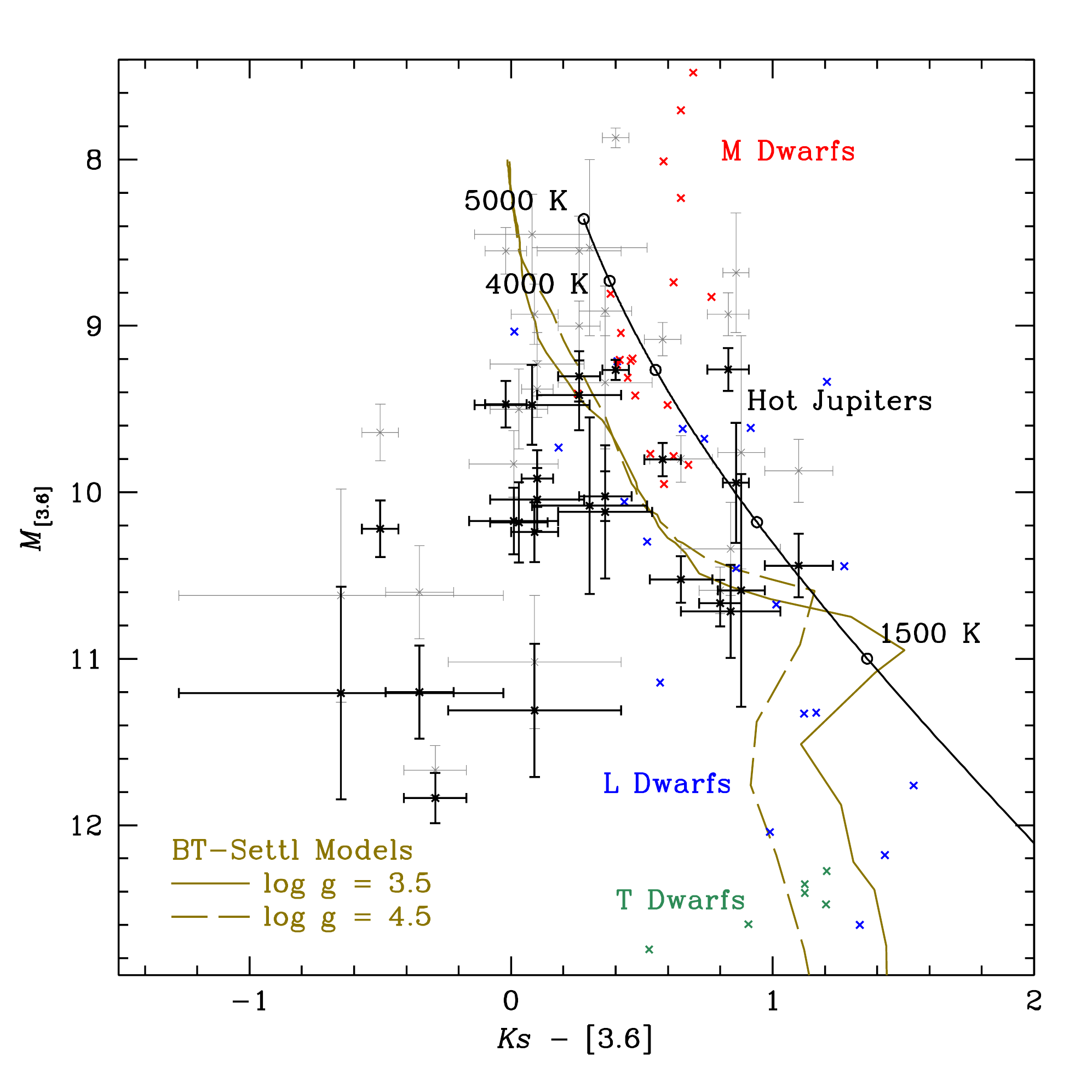}
     \includegraphics[width=8.8cm]{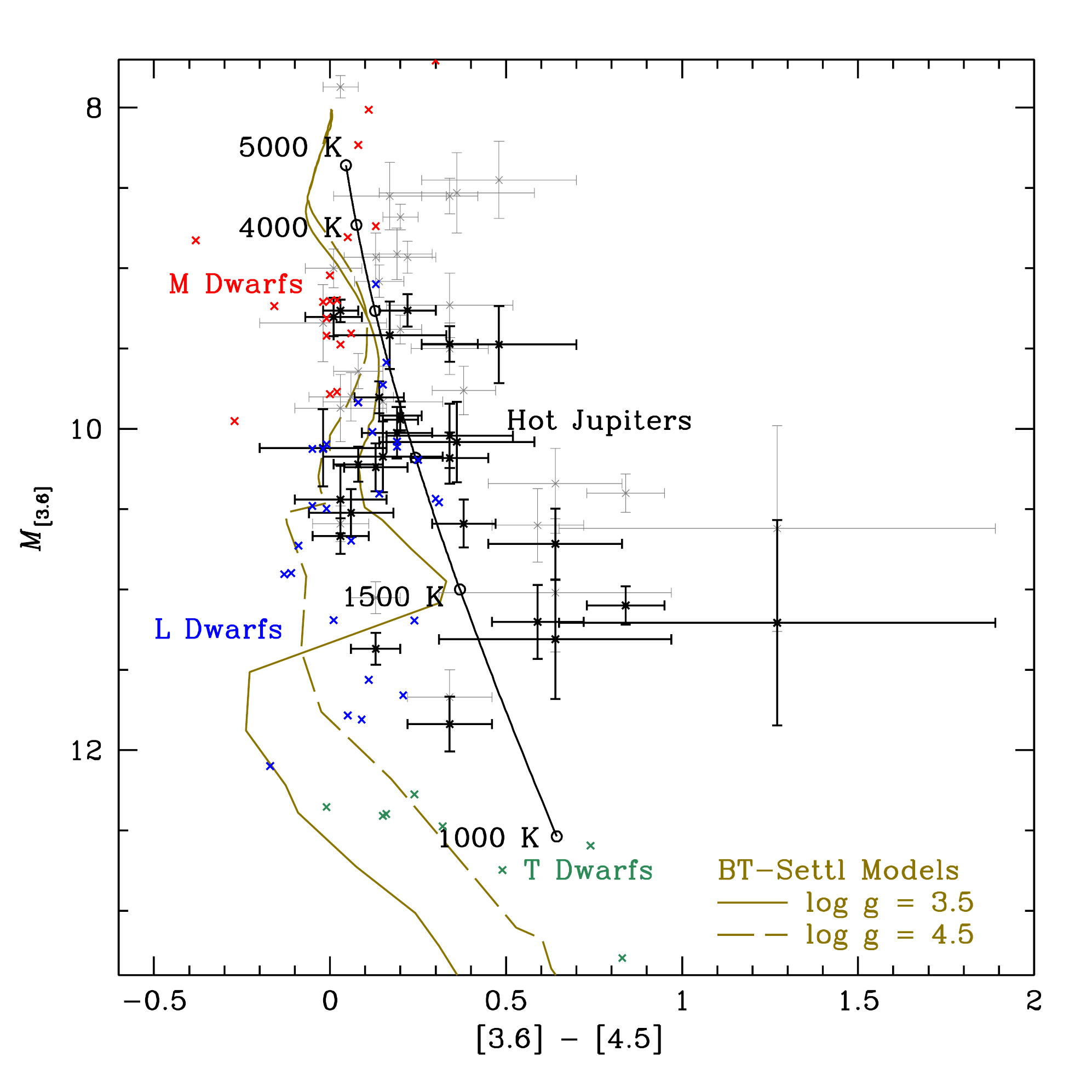}

    \caption{Colour-magnitude diagrams ($M_J$ vs J$-$H, $M_{K\!s}$ vs $J-K\!s$, $M_{[3.6]}$ vs $K\!s-$[3.6], $M_{[3.6]}$ vs [3.6]$-$[4.5]). Hot Jupiters are shown as grey points with error bars at the observed position and as black points with error bars at the position corrected for a radius of 1 $R_{\rm Jup}$. Red, blue and green crosses are ultracool dwarfs from the Database of Ultracool Parallaxes. Gold points are low-gravity dwarfs from \citet{faherty16}. The black body line for 1 $R_{\rm Jup}$ is shown as a solid black line with temperature labels. The BT-Settl models are shown as gold lines. See text for further details.}

    \label{fig:mag_col}
\end{figure*}

\section{Colour-Magnitude Diagrams}
\label{section4}

In Fig.~\ref{fig:mag_col} we show colour magnitude diagrams for hot Jupiters using the data described in section~\ref{section2}. We compare the hot Jupiter results with data for brown dwarfs, noting that these are mostly objects with higher mass but similar sizes and temperatures to hot Jupiters. Sub-stellar objects have radii largely independent of mass and so most brown dwarfs have radii not far from 1 $R_{\rm Jup}$ \citep{bailey14}. Highly irradiated hot Jupiters, however, have inflated radii \citep{baraffe10} up to around 2 $R_{\rm Jup}$ so are typically somewhat larger than the brown dwarfs despite their generally lower mass. On a colour magnitude diagram a change in radius results in a change in absolute magnitude and therefore a vertical shift on the diagram. To aid comparison with brown dwarf data and models we have therefore plotted each hot Jupiter data point on the diagram twice, once at its observed position (in grey) and secondly shifted vertically on the diagram (in black) to the position corresponding to a radius of 1 $R_{\rm Jup}$. The radii of these transiting planets are well determined and were taken from the Extrasolar Planets Encyclopedia (exoplanet.eu).

Red, blue and green crosses on the diagram are taken from the Database of Ultracool Parallaxes \citep{dupuy12,dupuy13,liu16}. Gold points are data for low gravity dwarfs taken from \citet{faherty16} that have been suggested as possible exoplanet analogs. The photometry from both these sources are plotted without error bars to avoid cluttering the diagram.

Also included on these diagrams are the black body line for an object of 1 $R_{\rm Jup}$ and the BT-Settl brown dwarf models \citep{allard12}. These are brown dwarf models including clouds that do a reasonable job of fitting the transition from L to T dwarfs that can be seen near the bottom of the $M_J$ vs $J-H$ and $M_{K\!s}$ vs $J-K\!s$ diagrams where the T dwarfs have bluer colours resulting in a sudden move to the left in the diagram away from the sequence shown by the M and L dwarfs. In brown dwarfs this transition is attributed to the disappearance of clouds that are important in the L dwarfs \citep{allard01,charnay18}.

The two upper panels of Fig.~\ref{fig:mag_col} show the $M_J$ vs $J-H$ and $M_{K\!s}$ vs $J-K\!s$ diagrams. In both of these the hot Jupiters are found to lie roughly along the black body line and in the same region as the M and L dwarfs, and the low gravity dwarfs from \citet{faherty16} which tend to lie a little to the right of the normal brown dwarfs on these diagrams. These low gravity dwarfs are suggested by \cite{faherty16} to be exoplanet analogs, and this seems consistent with what we see on these diagrams.

The bottom left panel showing $M_{[3.6]}$ vs $K\!s-$[3.6] presents a somewhat different picture with most of the hot Jupiters now lying to the left of the black-body line (i.e. bluer $K\!s-$[3.6] colours) and to the left of the brown dwarfs (although there are relatively few brown dwarfs with measured Spitzer colours that can be included on this plot). There are a few points with very blue colours (negative $K\!s-$[3.6]). Systematic errors in the magnitudes (the $K\!s$ magnitudes would have to be too bright, or the [3.6] magnitudes too faint) seems unlikely to explain this as it would lead to an opposite shift in the other diagrams involving these magnitudes which is not seen. However, it is possible that some of the objects with smaller eclipse depths are misplaced on this diagram due to overestimated $K\!s$ eclipse depths, as some of these objects do not have $J$ and $H$ measurements. Three of the objects that that lie at the bottom left of this diagram are WASP-10b, KELT-2Ab and HAT-P-1b all of which have only a single $K\!s$ band observation and among the lowest eclipse depths ($\sim$0.1\%).

At least the most extreme point, the leftmost one with the large error bars (which is WASP-46b) is probably due to a problem with the [3.6] eclipse depth measurement which has a large error and appears too low. This point deviates in the other direction on the $M_{[3.6]}$ vs [3.6]$-$[4.5] diagram. 

In the bottom right plot showing $M_{[3.6]}$ vs [3.6]$-$[4.5] the points once again scatter around the black body line.

\begin{figure*}
    \includegraphics[width=8.8cm]{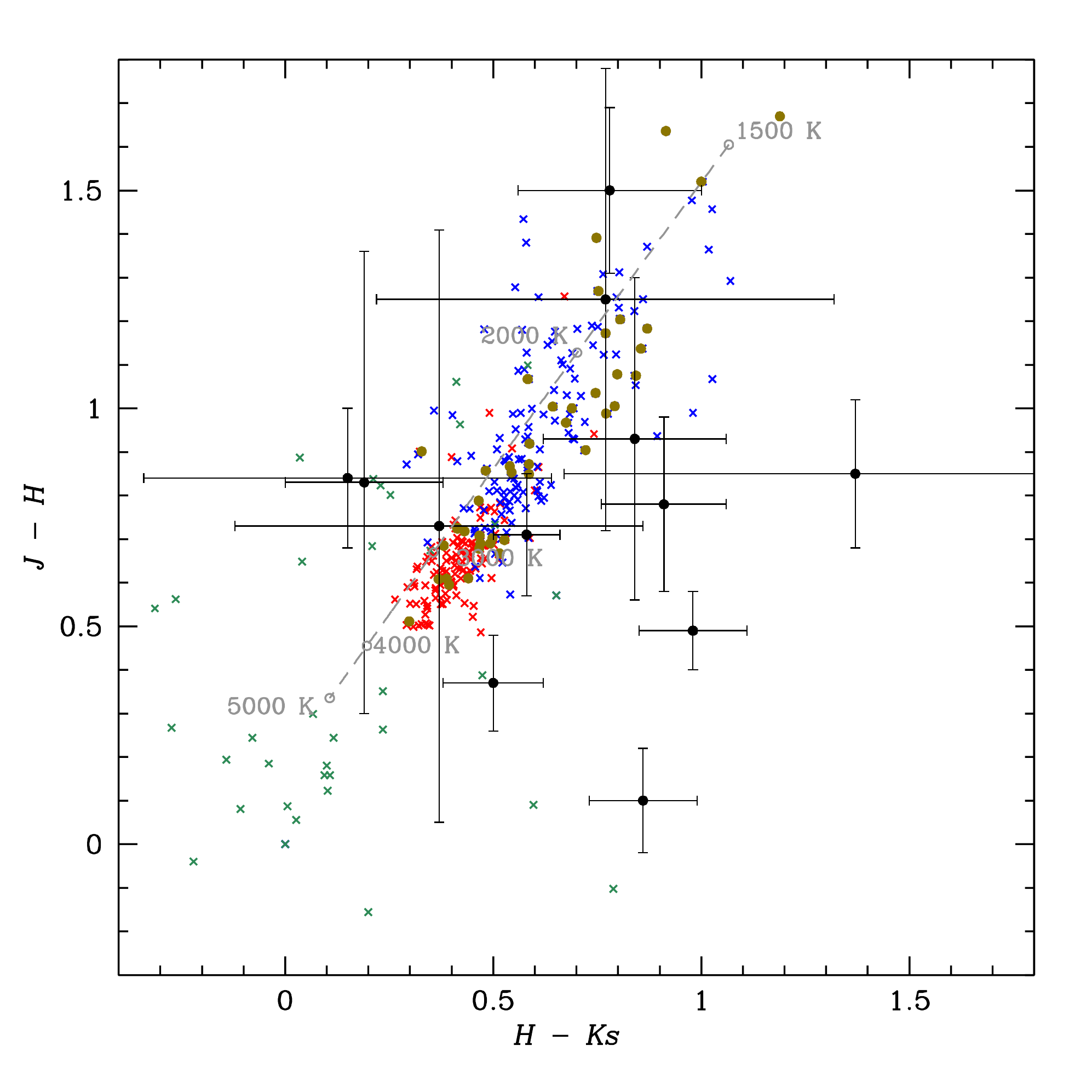}
     \includegraphics[width=8.8cm]{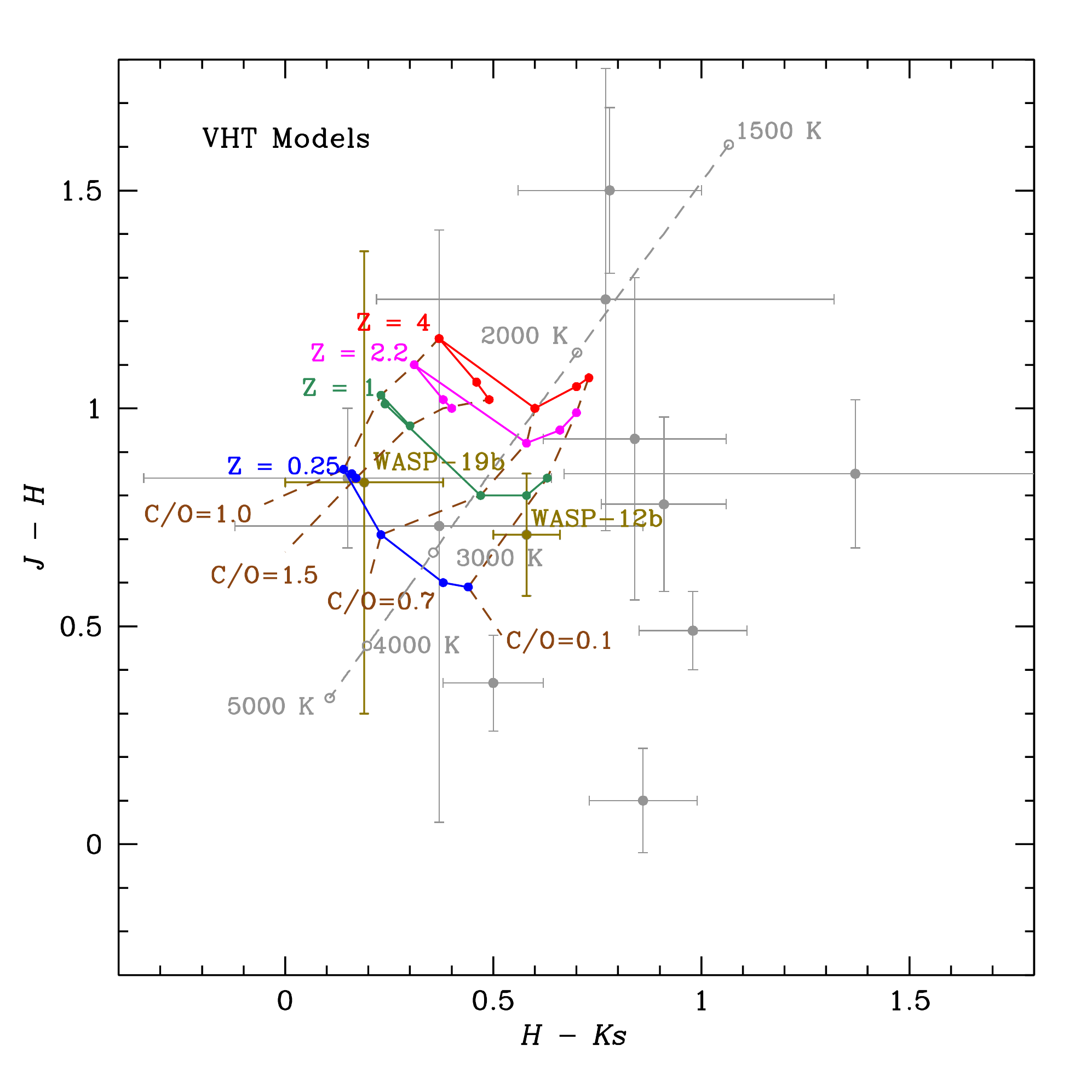}
     \includegraphics[width=8.8cm]{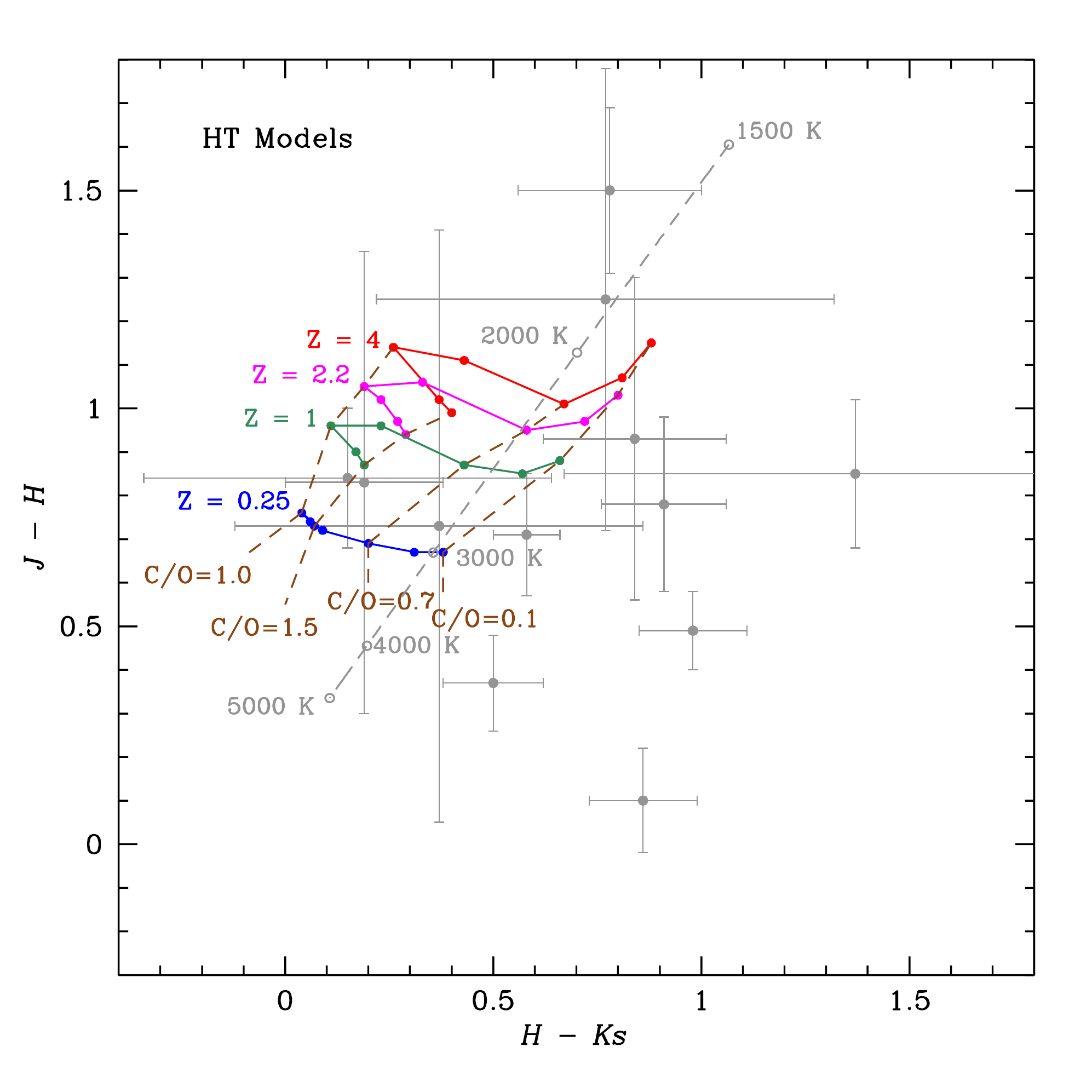}
      \includegraphics[width=8.8cm]{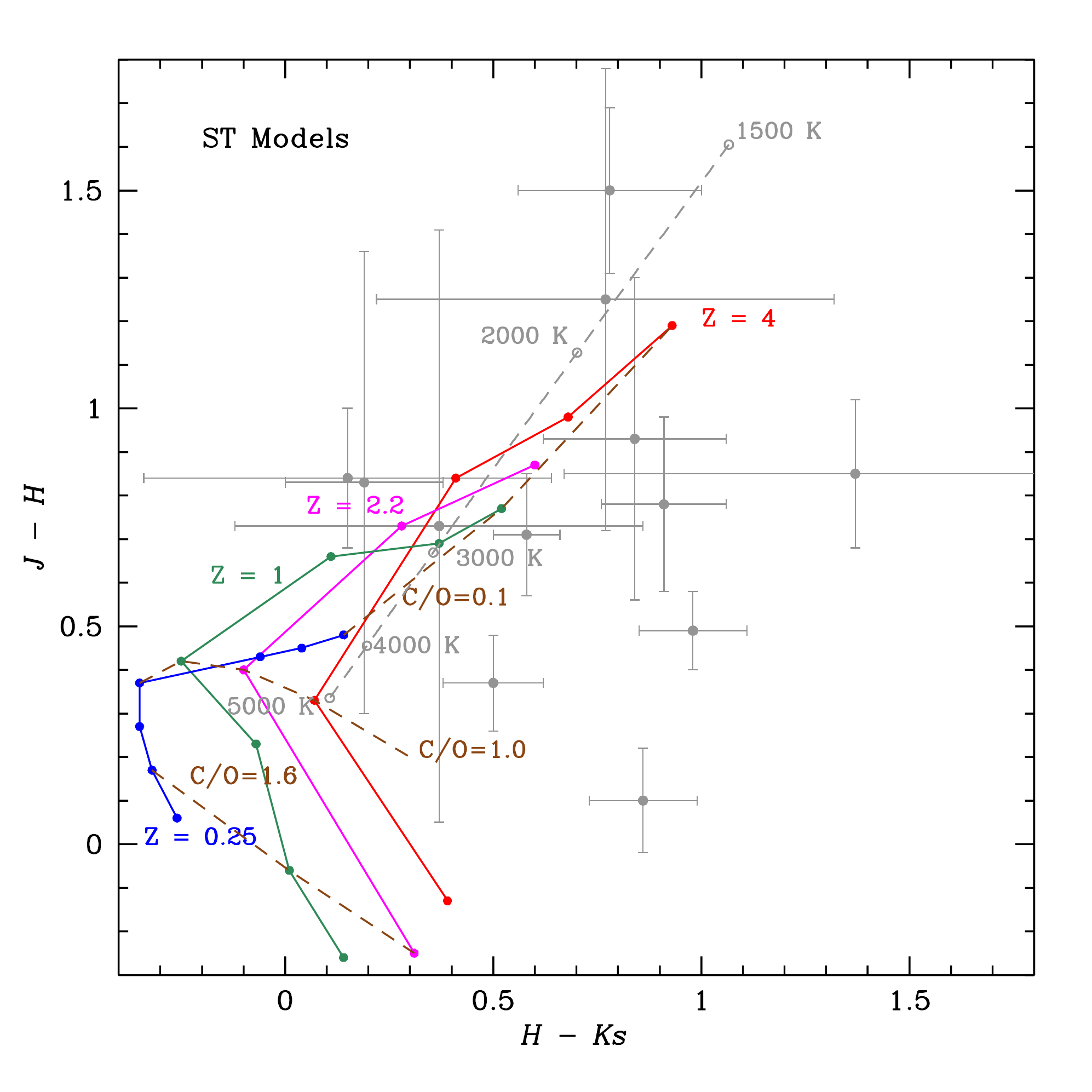}
      \label{fig:nr_col1}
    \caption{Colour-colour diagrams $H-Ks$ versus $J-H$ showing hot Jupiter observations compared to tracks showing model spectra with changing metallicity and C/O ratio. The top left panel shows the hot Jupiters compared with brown dwarfs and a black body line using the same symbols and colour codes as in Fig.~\ref{fig:mag_col}. Other panels show the data compared with the VHT, HT and ST models as described in section~\ref{section2}. The positions of WASP-19b and WASP-12b are also shown on the top right panel. 
    } 
\end{figure*}

\section{Colour-Colour Diagrams}
\label{section5}

In Fig.~\ref{fig:nr_col1} we present $H-K\!s$ versus $J-H$ diagrams that show tracks resulting from changes of metallicity and C/O ratio. For the VHT and HT models we see that with increased metallicity the $J-H$ colours become redder regardless of C/O ratio. Increasing the C/O ratio causes the $H-K\!s$ colours to become bluer though with a turn around in the sequence when C/O reaches 1.0. 

With sufficiently accurate observations these colours may be useful to set constraints on the atmospheric composition of hot Jupiters. However, it can be seen from Fig.~\ref{fig:nr_col1} that the error bars and scatter of the current hot Jupiter observations are such that it is not possible to put strong constraints on any individual object. However, in general the points cluster around the right hand end of the model tracks which is consistent with solar-like C/O ratios rather than the carbon rich values that are found at the left end. 

The lower temperature ST models show larger colour dependencies on metallicity and C/O but are outside the temperature range of the hot Jupiters for which $JHK$ colours are available. For ST models with C/O $>$ 1 we find relatively blue colours in both $J-H$ and $H-K\!s$ similar to the colours seen for T dwarfs (the green crosses in the top left panel of Fig.~\ref{fig:nr_col1}). The LT models, which are not plotted here, show even larger metallicity and C/O dependencies.

Combining data obtained in near- and in mid- infrared can be a potentially useful determinant of the atmosphere properties. We examined colour indices that involve at least one band measurement in each of these spectral regions and used $Ks$ band, since currently most measurements in near infrared are done in this band due to favourable brightness of objects and a possibility of water absorption detection. We tested different combinations of such colours and found that some of them are more useful than others, because different parameters led to very distinct colour indices.

\begin{figure*}
    \centering
    \includegraphics[width=8.8cm]{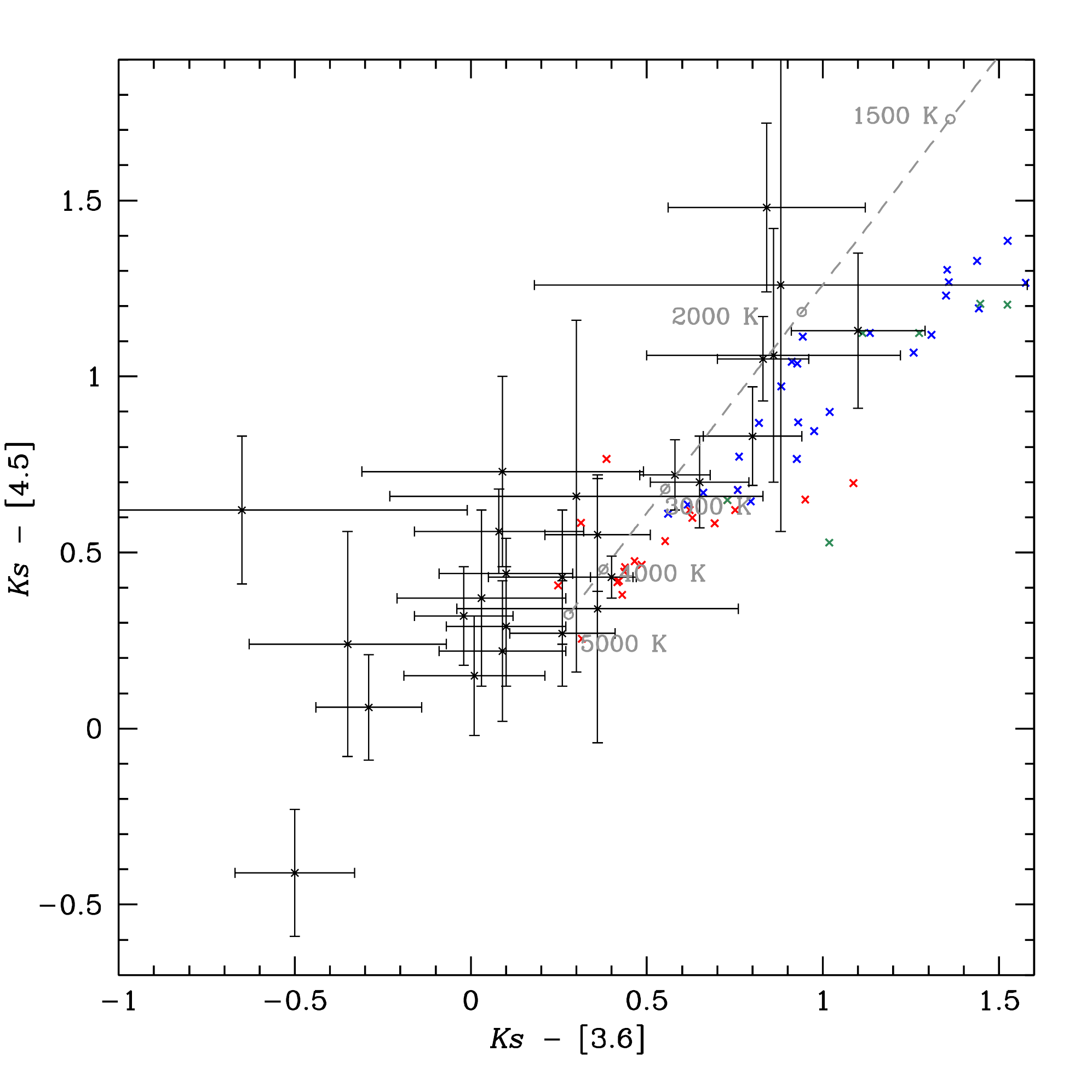}
    \includegraphics[width=8.8cm]{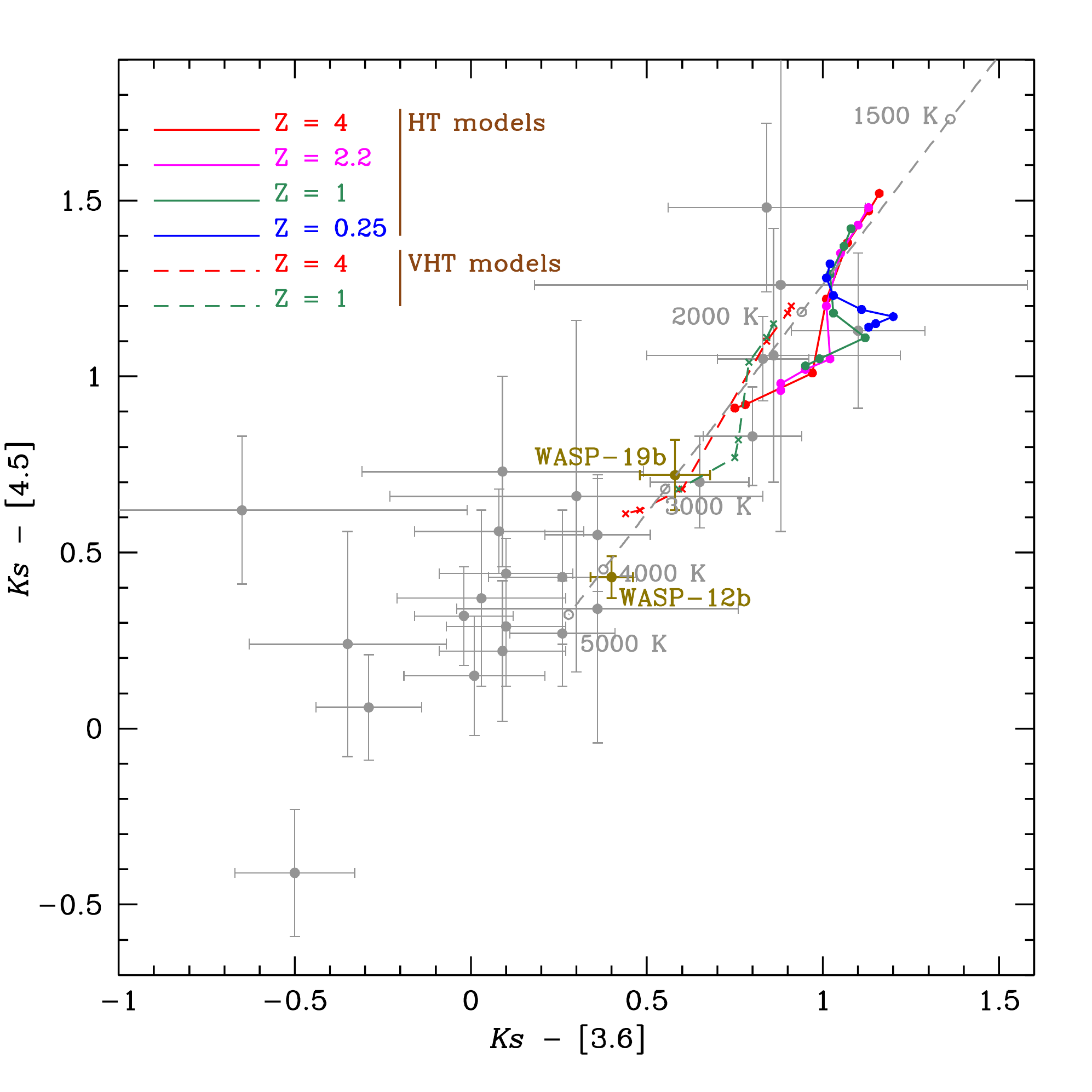}
    \caption{Colour-colour diagram of $K\!s-$[3.6] versus $K\!s-$[4.5]. The left panel shows the position of a number of observed hot Jupiters with error bars compared with brown dwarfs using the same symbols and colours as Fig.~\ref{fig:mag_col}. The right panel shows the comparison with the HT and VHT models. The positions of WASP-19b and WASP-12b are also shown on the right panel.} 
    \label{fig:sp_col1}
\end{figure*}

In Fig.~\ref{fig:sp_col1} we present $K\!s-$[3.6] versus $K\!s-$[4.5] colours. The left panel shows a number of observed hot Jupiters along with brown dwarfs. It indicates that many of the observed hot Jupiters tend to lie near the blackbody line. However as was seen in the colour magnitude diagrams many hot Jupiters have blue $K\!s-$[3.6] colours putting them at the left of this diagram in a region bluer than any of the brown dwarfs and also bluer than our model predictions for the HT and VHT models (seen in the right panel). As discussed in section~\ref{section4} some of these cases may be due to overestimated $K\!s$ eclipse depths, but this seems unlikely to explain all these cases. Alternatively the placement of these objects may indicate that our models do not correctly represent the temperature pressure profiles needed for these objects. For example they may have temperature inversions which are not considered in the models used here.

In this diagram models of different metallicity and C/O ratio all lie close to the black body line, and thus provide even less discrimination of composition than was seen in the near-infrared colours.

\section{Discussion}
\label{section6}

We find that hot Jupiters occur in the same region of the $JHK\!s$ colour magnitude and colour colour diagrams as brown dwarfs. These brown dwarfs share many characteristics with planets including a range of temperatures which leads to a change in their atmospheres from M-dwarf like spectra to planet-like spectra. Brown dwarfs atmospheres are heated from below and as such have a PT profile that steadily decreases in temperature with decreasing pressure. Hot Jupiter atmospheres, however, have a PT profile where the temperature decreases up to around 1 bar and then remains fairly steady, as in Fig.~\ref{fig:pt}. Late M and L Dwarfs have dusty atmospheres \citep{allard01}  which tend to suppress spectral features and cause them to cluster around the black body line. However, the cooler T Dwarfs have largely clear atmospheres and are found to have bluer colours away from the black body line (Fig.~\ref{fig:mag_col}  and Fig.~\ref{fig:nr_col1}). Their spectra are dominated by absorption bands from methane and water, and have very broad absorption features from the alkali metals Na and K.  They also lack the FeH and CrH bands that L type dwarfs exhibit.   Another contributing factor affecting all brown dwarf atmospheric profiles in general, is their higher gravity. This is due to their larger compact mass then hot Jupiters which leads to a different atmospheric scale height. 

Although hot Jupiters are found in the same region of the diagrams as these dusty brown dwarfs, it seems unlikely that they are there for the same reason. Many hot Jupiters are simply too hot for dust to condense (e.g. WASP-18b, \citet{chudczer19}). Many hot Jupiter's have very low geometric albedos as seen in Kepler secondary eclipse measurements \citep{bailey14} implying a lack of reflecting clouds, although there are exceptions such as Kepler-7b that do show clouds \citep{demory13}. Analysis of transit spectra show a range from clear to cloudy atmospheres \citep{sing16}. Thus while some hot Jupiters have clouds they do not have the uniformly dusty atmospheres found in brown dwarfs of the same temperature range.

In the case of hot Jupiters the reason for the black body like colours is more likely to be the steeper temperature variation with pressure that arises in an externally heated atmosphere, and leads to a near constant temperature over a wide range of pressure. This has the effect of suppressing spectral features and leading to a black body like spectrum. The agreement between the colours of brown dwarfs and hot Jupiters, is because both, for different reasons, are behaving like black bodies.

The models presented in this paper represent a fraction of the total number of cases analysed, although the key ones are shown here. Diagrams have been constructed for objects over a wide range of temperatures as well as C/O ratio and metallicity. This seems particularly needed in the current situation where the large number of discovered hot-Jupiters are indeed in the hottest range of models considered here. However some of these hottest planets are likely to have more complex atmospheric profiles that include temperature inversions due to heat trapping molecules of TiO and VO, and could not be uniformly compared in this study. 

 We also ran some simulations to look at how a change in gravity could affect the colours of a planet and its position on the various diagrams such as the $K\!s-$[3.6] versus $K\!s-$[4.5] diagrams. We found small there were small shifts in position but they were not substantial. We ran simulations for [HT] profiles with surface gravity lower than, 24.8 ms$^{-2}$, of a Jupiter-like planet, such as might be encountered in a more inflated planet.  For constant metallicity and C/O ratio, decreasing surface gravity makes both colour indices redder. This behaviour is consistent for higher C/O ratios and metallicity range tested in our models. When we compare this trend to the variation in C/O ratio for the single value of Jupiter surface gravity it appears that both tracks are almost overlapping leading to potential degeneracy between both parameters. This degeneracy can be lifted when the data is examined at the diagram that involves only near infrared colours where the variable C/O sits at a lower $J-H$ and $H-Ks$. It is worth noting that the effect of shifting the colour indices towards red is of the order of 0.2, that is, below the typical error bar on the current measurements of magnitudes for hot-Jupiters. 
 
To look at some specific cases of well-studied hot Jupiters we consider the cases of WASP-12b and WASP-19b. WASP-19b is a hot-Jupiter that has published measurements in all bands and was previously modelled with an atmospheric profile similar to our [VHT] profile \citep{2013ApJ...774..118Z}. That study concluded that models with a high C/O ratio best fit the data. We show the position of WASP-19b on the top-right panel of Fig.~\ref{fig:nr_col1}. On this figure it does lie near the region of the VHT models for high C/O ratio. However, the error bars are large and consistent with a wide range of compositions. When the object is placed on the $K\!s-$[3.6] versus $K\!s-$[4.5] diagram (Fig.~\ref{fig:sp_col1}) it again falls into the region of [VHT] models, but it is not possible to say anything siginificant about its C/O ratio from its position in this diagram. WASP-12b is another hot Jupiter that has been suggested as having a high C/O ratio on the basis of its dayside emission spectrum \citep{2011Natur.469...64M}. It can be seen however that on the $J-H$ Vs $H-K\!s$ diagram (Fig.~\ref{fig:nr_col1}) it falls on the region of VHT models for low C/O ratio, and on the $K\!s-$[3.6] versus $K\!s-$[4.5] diagram it is part of the large group of objects that lie at bluer colours than any of the models. It can be seen that no strong conclusions about the composition of these atmospheres can be drawn on the basis of these diagram.
The detection of H$_{2}$O absorption in WASP-19b (Oshagh, 2017) indicates that the C/O ratio is probably less than one. Generated spectra from all models could also be valuable and compared to future spectra from the JWST.

A number of simulations show a sharp change in trend direction on the C/O diagrams such as Fig.~\ref{fig:nr_col1}. This sharp trend change, as mentioned earlier, is due to the variation in the C/O ratio and shows most strongly around C/O$=1$ where the dominance of absorption from oxygen-rich species is being replaced with increasing contribution of carbon-rich species. This can also be seen in  Fig.~\ref{fig:spectra}.  At the lowest temperatures, the horizontal V-shape, Fig.~\ref{fig:nr_col1}, becomes more prominent at higher metallicities. This could possibly be caused by the widening and deepening methane absorption band at 3.3$\mu$m. This has been stated to occur in brown dwarfs' atmospheres \citep[e.g.][]{2006ApJ...651..502P} and leads to increasingly redder colours with increasing magnitudes.

Although the simulations in this paper are based on specific ranges in temperature, metallicity, C/O ratio and mostly a fixed gravity value, they are not unreasonable values considering the available data on these parameters from observational studies and theoretical work. That is, these parameter values are expected to cover the vast majority of hot Jupiters. In addition, although the modelling assumes no prominent atmospheric inversions, there has been growing evidence that temperature inversions are nowhere near as common as previously thought. As for the likelihood of dust, aerosols, clouds or haze, research is continuing.

\section{Conclusions}
This paper has presented colour-magnitude and colour-colour diagrams for hot Jupiters
using near infrared ($JHK\!s$) and Spitzer 3.6 and 4.5 $\mu$m photometry. The number of objects available for such analysis is much increased since the previous studies of \citet{triaud14a}, \citet{triaud14b} and \citet{zhou15}, particularly in the near infrared bands. The availability of GAIA DR2 parallaxes for most objects also increases the accuracy of the absolute magnitudes.

On colour magnitude diagrams involving the $JHK\!s$ magnitudes, hot Jupiters are found to lie close to the black-body line and in the same region as late M and L dwarfs, agreeing well with the low gravity dwarfs suggested to be exoplanet analogs by \citet{faherty16}. We suggest the reasons for this location may be different in brown dwarfs and hot Jupiters. For brown dwarfs it is well established that dust clouds are important in these systems, and it is the presence of this dust that causes the flux distribution to approach that of a black body. In hot Jupiters, where clouds are probably less important, as indicated by the low albedos, it is the approximately isothermal temperature profile (such as those in Fig.~\ref{fig:pt}) which causes the spectrum to be similar to a black body.

Using a set of models with different metallicities and C/O ratios we investigate whether these diagrams can help constrain the atmospheric composition of hot Jupiters. We find that there are systematic trends with metallicity and C/O particularly on the shorter wavelength diagrams such as $H-K\!s$ Vs $J-H$. However, the error bars on current observations at these wavelengths are such that we cannot make conclusive statements about the composition of individual objects. As a class, however, hot Jupiters cluster around the colours expected for solar metallicity and C/O on this diagram. The trends with metallicity and C/O become larger for lower temperature objects (our ST and LT models), but the currently available eclipse observations do not extend to planets as cool as this. 

Diagrams involving the longer wavelength colours such as $K\!s-$[3.6] Vs $K\!s-$[4.5] are less helpful for constraining composition since the effects of metallicity and C/O are smaller, but also because many observed hot Jupiters have bluer $K\!s-$[3.6] colours than any of the models. This might indicate that some systems have overestimated $K\!s$ eclipse depths, or that the temperature-pressure profiles of our models do not properly represent these objects.

At present there is very limited infrared spectral data on hot Jupiters due to observational challenges. The observational data to come in the following years, especially from the James Webb Space Telescope (JWST), will undoubtedly test present hot Jupiter models. The large error bars in the IR bands of observational ‘hot Jupiter’ exoplanets, which make comparative work on temperature, metallicity and C/O using our models very difficult, will be overcome by the JWST, which will provide full spectra. The JWST will open a new phase in our understanding of exoplanets with transit spectroscopy of relatively short period planets and coronagraphic imaging of ones with wide separations from their host stars \citep{beichman19}. Of course, the complex chemistry at play further complicates things but as we build up more information on what is happening in hot-Jupiter atmospheres, over time, the many pieces of the jigsaw will slowly fall into place.  
\section*{Acknowledgements}

\label{sec:acknowledgements}
Based in part on data acquired through the Australian Astronomical Observatory. We acknowledge the traditional owners of the land on which the AAT stands, the Gamilaraay people, and pay our respects to elders past and present.
We thank Dr George Zhou for discussions and ideas that improved this work.

This work has made use of data from the European Space Agency (ESA) mission
{\it Gaia} (\url{https://www.cosmos.esa.int/gaia}), processed by the {\it Gaia}
Data Processing and Analysis Consortium (DPAC,
\url{https://www.cosmos.esa.int/web/gaia/dpac/consortium}). Funding for the DPAC
has been provided by national institutions, in particular the institutions
participating in the {\it Gaia} Multilateral Agreement.




\bibliographystyle{mnras}
\bibliography{mybib1} 








\bsp	
\label{lastpage}
\end{document}